\documentstyle[amssymb,preprint,aps]{revtex}

\begin{document}
\title{Influence of Nd on the magnetic properties of $Nd_{1-x}Ca_{x}MnO_{3}$: an
ESR study}
\author{F.\ Dupont $^{1}$, F. Millange$^{2}$, S.\ de Brion$^{1\ast }$, A. Janossy$%
^{1,3}$, G. Chouteau$^{1}$}
\address{$^{1}$Grenoble High Magnetic Field Laboratory\\
BP 166 - 38042 Grenoble cedex 9- France\\
$^{2}$ CRISMAT-ISMRA\\
CNRS\ and Universit\'{e} de Caen\\
6, boulevard du Mar\'{e}chal Juin\\
14050 Caen - France\\
present address: Institut Lavoisier\\
Universit\'{e} de Versailles-Saint-Quentin-en-Yvelines\\
45, Avenue des Etats-Unis, 78035 Versailles cedex, France.\\
$^{3}$ Budapest University of Technology and Economics, Institut of Physics,%
\\
H 1521 , POB 91, Hungary\\
$^{\ast }$ corresponding author}
\maketitle
\date{\today }

\begin{abstract}
The role played by the Nd ions in the magnetic properties of $%
Nd_{0.5}Ca_{0.5}MnO_{3}$ and $Nd_{0.7}Ca_{0.3}MnO_{3}$ is studied using
static magnetization, neutron diffraction and high frequency $(9.4-475GHz)$
Electron Spin Resonance. We show that the Nd ions are weakly coupled to the
Mn ions via ferromagnetic exchange and are responsible for the peculiar
ferromagnetic resonance observed in the FM\ phase of both compounds (ground
state below 120K for x=0.3, high field state for x=0.5).\ We then use ESR to
look for magnetic phase separation in the low field, CO phase of $%
Nd_{0.5}Ca0.5MnO_{3}$. We show that there is {\it no trace\ of the FM\ phase}
imbedded in the CO\ phase, contrary to what is observed in $%
La_{0.5}Ca_{0.5}MnO_{3}$ or $\Pr_{0.5}Sr_{0.5}MnO_{3}$.
\end{abstract}

.

\newpage

The perovskite manganites, $R_{1-x}A_{x}MnO_{3}\,$, where R is a trivalent
rare earth element (La, Pr, Nd, Sm etc.) and A is a divalent alkaline earth
element (Ca, Sr), are strongly correlated electron systems distinguished by
their large variety of magnetic and electronic states. Three types of order
are observed: charge order (CO) of the $Mn^{3+}$and $Mn^{4+}$ charges;
orbital order (OO) of the manganese 3d orbitals and antiferromagnetic (AF)
or ferromagnetic (FM) order of the magnetic moments \cite{Review}. The
various types of order are related, e.g. the onset of orbital order induces
an antiferromagnetic order with the same symmetry. On the other hand, the
ferromagnetic double exchange interaction between $Mn^{3+}$and $Mn^{4+}$
ions leads to a metallic ferromagnetic state with charge delocalization and
no orbital order. The type of order and the ordering temperature may be
tuned by the choice and concentrations of the doping elements $R^{3+}$ and $%
A^{2+}$; these affect the bonding length, angle and disorder of the Mn-O-Mn
overlapping orbitals and the relative concentration of $Mn^{3+}$and $Mn^{4+}$
ions. A sufficiently large magnetic field may suppress the OO or the CO and
stabilize a FM state. In the Ca compounds, the charge ordered state is quite
stable : there is no delocalized state with a FM order at any temperature in
zero magnetic field. In the typical examples, $Nd_{0.5}Ca_{0.5}MnO_{3}$ and $%
Pr_{0.5}Ca_{0.5}MnO_{3}$, there is a transition to a CO state at $250K$ and
an antiferromagnetic state at $160K$ \cite{PrCa}, \cite{PRBNdCa}. A high
magnetic field destroys the CO state and stabilizes the FM delocalized state.

The strength of spatial and dynamical fluctuations of the order in
paramagnetic (PM), CO or AF manganites balanced close to the delocalization
- localization transition is one of the much debated questions. Allodi {\it %
et al} \cite{Allodi} observed, using NMR, a separation between ferromagnetic
and antiferromagnetic phases in $La_{0.5}Ca_{0.5}MnO_{3}$ and $%
\Pr_{0.5}Sr_{0.5}MnO_{3}$. The coexistence of the charge ordered and
ferromagnetic metallic phases has been observed by electron microscopy in
other $R_{1-x}Ca_{x}MnO_{3}$ compounds also \cite{Uehara}.

In this Rapid Communication, we study the $Nd_{1-x}Ca_{x}MnO_{3}$ manganites
belonging to the Ca family. We present magnetization and electron spin
resonance (ESR) data together with neutron diffraction determination of the
Mn and Nd moments for the $x=0.3$ ($Nd_{0.7}Ca_{0.3}MnO_{3}$) and $x=0.5$ ($%
Nd_{0.5}Ca_{0.5}MnO_{3}$) compounds. We show that the magnetic moment of the
trivalent Nd ion has little influence on the magnetic phase diagram of the
system as it is only weakly coupled by a ferromagnetic exchange to the Mn
ions. The magnetic resonance spectra are, however, sensitive to the Nd
moment that is increased by the coupling to the Mn sublattice. Once the role
of Nd on the spectrum is understood, we use ESR to look for magnetic phase
separation. We show that, there are no static FM domains in the low field CO
state of the $Nd_{0.5}Ca_{0.5}MnO_{3}$ compound.

Preparation of the powder samples is described elsewhere \cite{PRBNdCa}, 
\cite{Franck}. The neutron diffraction data presented here were performed at
ILL and are part of a detailed study to determine the crystallographic
structure, charge order and magnetic transitions in zero field. The magnetic
phase diagrams of our samples constructed from static magnetization data in
fields up to $23T$ are in agreement with those of Tokunaga {\it et al} \cite
{phase diagram NdCa}. The X band $(9.4GHz)$ and Q band $(35GHz)$ ESR spectra
were recorded with conventional Brucker spectrometers on few mg loosely
packed powder samples with $7\mu m$ typical grain size. A home made
spectrometer \cite{AMRfred} was used at higher frequencies $(95-475GHz)$
with a superconducting magnet up to $12T$ and a resistive magnet for higher
fields. The sample was pressed into $9mm$ diameter, $0.9mm$ thick pellets
with a density of $82\%$ for the high field ESR measurements. The magnetic
field was oriented along the axis of the cylindrical pellet. In this
geometry, the demagnetizing fields shift the resonance to higher fields by $%
\delta H_{dem}\simeq \mu _{0}M$ and the shift data were corrected
accordingly using the static magnetization, $M$, measured on the same sample.

The ground states in zero magnetic field are FM and AF for the $x=0.3$ and $%
x=0.5$ compounds respectively. In the $x=0.3$ compound, the paramagnetic CO
state orders below $T_{CO}=240K$ into a FM state at $T_{C}=120K$. In the $%
x=0.5$ compound, the transition to the paramagnetic CO state at $T_{CO}=250K$
is well above the AF ordering temperature, $T_{N}=160K$ in zero magnetic
field. A magnetic field of about $15T$ is required to stabilize the FM state
below $275K$. In this high field, the CO state is entirely suppressed.

Fig.1 shows the static magnetic susceptibility, $\chi _{DC}$, determined
from the magnetization measured in a field of $0.33T$. The susceptibility
measured by the ESR intensity follows $\chi _{DC}$ in the PM state of both
the $0.3$ and $0.5$ compounds. The maximum observed in $\chi _{DC}$ for $%
x=0.5$ at the CO transition is confirmed. For $x=0.3$, $\chi _{DC}$
increases continuously as the temperature decreases from the paramagnetic
state to $75K$, i.e. well into the FM state. The ordered magnetic moments of
the Mn and Nd ions measured by neutron diffraction in zero field are shown
on Fig.2. The Mn moments order below $120K$ while the Nd contribution is
visible only below $50K$. The Nd moment is still increasing between $10K$
and $2K$. A similar neutron diffraction study in the AF state of the $x=0.5$
compound showed no magnetic order of the Nd subsystem. We relate the upturn
in $\chi _{DC}$ below $25K$ in the antiferromagnetic state of $x=0.5$
compound (Fig.1b) to the increase of the paramagnetic Nd moment within the
antiferromagnetically ordered Mn system. Thus, both the static
susceptibility and neutron data confirm that the coupling of Nd to the Mn
ordered lattice is weak, and much weaker in the AF than in the FM state.

The ESR data discussed below show that the coupling between Nd and Mn is
ferromagnetic and is well observable both in the FM and PM states. Above $%
T_{C}=120K$, i.e. in the paramagnetic state of the $x=0.3$ compound, a
single symmetric ESR line is observed. The spectrum broadens below the FM
ordering temperature; as the temperature is lowered a double peaked
structure develops that shifts to lower fields (Fig3a). The shift is far
larger and in opposite direction than the corrections due to demagnetization
effects. We observe a similar ESR spectra at $475GHz$ in the high field FM
state of the $x=0.5$ compound (Fig3b). The single line centered around $17T$
in the PM state becomes double peaked in the FM state below $T_{C}=275K$. At
low temperatures, the demagnetization corrected shifts of both components
are large and negative. We note that the unusual features of the resonance,
the double peaked shape and large shift, are specific to manganites
containing Nd. In $La_{0.67}Ca_{0.33}MnO_{3}$, a compound that is
ferromagnetic below $280K$, the ESR of an epitaxial film showed no unusual
feature\cite{LaCafilm}. In this case the spectrum in the ferromagnetic phase
could be described by demagnetization effects and a relatively small
magnetic anisotropy. Long range structural or magnetic phase inhomogeneities
may lead to spectra with several components as observed in $%
La_{1-x}Ca_{x}MnO_{3}$ \cite{LaCadoublespectra} and $%
La_{1.35}Sr_{1.65}Mn_{2}O_{7}$ \cite{Chauvet} but in this case the
temperature dependence of the various components are unrelated.

The large shift with temperature of all components of the spectra in the FM
state of $Nd_{1-x}Ca_{x}MnO_{3}$ for both $x=0.3$ and $0.5$ indicate that
these are phase homogeneous compounds. We suggest that the spectra below $%
T_{C}$ are the ferromagnetic resonance of the ferromagnetically ordered Mn
moments weakly coupled by a ferromagnetic exchange to the paramagnetic Nd
moments. As discussed above the neutron diffraction and magnetization data
show that Nd moments are weakly coupled to the Mn ferromagnetic system. We
assume that this coupling is so weak that Nd and Mn do not have a common
resonance. This will happen if the spin lattice relaxation of Nd due to e.g.
phonons is sufficiently rapid so that there is no bottleneck between the Mn
and Nd subsystems. The spin relaxation and g factor anisotropy broadens the
Nd ESR beyond observability. The shift of the Mn ferromagnetic resonance is
then due to the exchange interaction between Nd and Mn: $\delta
H_{Mn}=\lambda M^{Nd}$. The observed anisotropy for $\delta H_{Mn}$ results
from the Nd g anisotropy.

In Fig.2 the average shift of the Mn ESR line is compared to the temperature
dependence of the Nd moment measured in zero field. In the ESR experiment
the polarization of Nd arises from the external field, the ferromagnetic
interaction with the ordered Mn sublattice and the ferromagnetic interaction
between Nd moments. Inter-neodymium interactions are small above 50 K.\ In
the FM state the polarization of Nd by the external field is smaller than
the polarization due to the Mn sublattice.\ Indeed, at $10K$ the shift (thus
the Nd moment) increases little $(8\%)$ between $95GHz$ and $285GHz$
(roughly $2T$ and $9T$ central fields), at $60K$ this increase is much more
important $(140\%)$. Thus at $10K$ the extrapolation to zero external field
and the moment of $0.9\mu _{B}$ per Nd ion measured by neutron diffraction
yields the scale for the larger (smaller) Mn shift of $\lambda =2.4T/\mu
_{B}Nd$ $(1.3T/\mu _{B}Nd)$. We extract the Nd susceptibility in the whole
temperature range $10K-300K$ (Fig.1a) using the average value of $\lambda
=1.8T/\mu _{B}Nd$ . The same analysis can be applied to the high field FM\
phase of $Nd_{0.5}Ca_{0.5}MnO_{3}$. At $50K$, the same Nd g anisotropy is
found.\ Taking the same value for $\lambda $ for both compounds ( $x=0.3$
and $x=0.5$), we estimate the Nd susceptibility (Fig1.b). In both samples
the Nd susceptibility increases faster than $1/T$ at low T while the Mn
moments are nearly saturated. This shows that Nd orders ferromagnetically at
much lower temperature than Mn\ and confirms that the coupling of the Nd to
the Mn ordered lattice is weak, and much weaker in the AF\ state than in the
FM state.

We examine now the low field, paramagnetic CO phase of $%
Nd_{0.5}Ca_{0.5}MnO_{3}$. The above discussion shows clearly the kind of
spectrum expected from phase separated FM domains within the PM state: if
long range FM domains existed in this material then these would be detected
as separate lines or at least as an increased tail at the low field side of
the spectra since the Nd moments shift the Mn ESR strongly to lower fields.
Fig.4 presents typical spectra taken in the CO\ phase of $%
Nd_{0.5}Ca_{0.5}MnO_{3}$ as well as the temperature dependence of the line
position. The susceptibility measured by ESR follows the general trend of
the static magnetic susceptibility (Fig.1b), the difference is due to the Nd
contribution which is not present in the ESR\ susceptibility.\ It is clear
that the ESR spectra, in the CO phase at low fields, are simpler than in the
high field induced FM phase. The line remains symmetric and centered around $%
g=1.99$ in the whole temperature range between $160K$ and $300K$. There is
no shift of the resonance line contrary to what is observed in the FM phase.
As shown in the inset of Fig. 4b, no extra peaks or asymmetry in the
lineshape are observed. \ Besides, at $230K$, the line width in the CO phase
is smaller than the total spread of the ferromagnetic spectrum in the FM
state. Therefore we conclude that {\it there is no trace of FM domains
embedded in the CO matrix}. The scenario of phase separation does not hold,
contrary to what is observed in $La_{0.5}Ca_{0.5}MnO_{3}$ or $%
\Pr_{0.5}Sr_{0.5}MnO_{3}$ \cite{Allodi}. In those compounds, the
ferromagnetic phase is more easily stabilized by an applied magnetic field;
the energy difference between the CO and FM phases is reduced and the phase
separation scenario may be valid: the FM phase may nucleate into the CO
phase. In Nd compounds, as well as in Pr compounds, the energy difference is
larger and the nucleation of the ferromagnetic phase occurs only at high
fields, closer to the CO-F transition. The stability of the CO phase is
quite robust against changes in the $Mn^{3+}/Mn^{4+}$ concentration: at $%
x=0.4$, we observe a similar behavior as for $x=0.5$. It is only at $x=0.3$
that a ferromagnetic state prevails in zero magnetic field.\ 

We thank P.Monod for helpful discussions on the ESR results. We also
acknowledge A.\ Maignan and C. Simon for their interest in this work. The
Grenoble High Magnetic Field Laboratory is 'laboratoire conventionn\'{e}
\`{a} l'Universit\'{e} Joseph Fourier'. A.J.\ acknowledges funding from the
Hungarian Science Fund OTKA T029150.

\section{Figure captions}

{\bf Figure 1}: temperature dependence of the DC magnetic susceptibility ($%
\blacksquare $), ESR manganese\ susceptibility\ ($\Delta $) normalized at
room temperature and Nd susceptibility ($\square $)calculated from ESR line
shifts at $0.33T$ for $Nd_{1-x}Ca_{x}MnO_{3}$, $x=0.3$ (Fig.1a) and $x=0.5$
(Fig.1b). The magnetic and charge order transitions have been determined by
neutron diffraction at zero field.

{\bf Figure 2}: temperature dependence of the Mn ($\blacksquare $)\ and Nd ($%
\blacktriangle $) ordered ferromagnetic moment determined by neutron
diffraction as well as the average ESR line shift ($\bigcirc $) at $95GHz$
for $Nd_{0.7}Ca_{0.3}MnO_{3}$.

{\bf Figure 3: }ESR\ absorption line position (corrected from the sample
demagnetization field) as a function of temperature in $%
Nd_{0.7}Ca_{0.3}MnO_{3}$ taken at $95GHz$ (a) and in $%
Nd_{0.5}Ca_{0.5}MnO_{3} $ FM phase taken at $475GHz$ (b). the dotted line
corresponds to $g=1.99$. Insert: ESR absorption spectrum at different
temperatures.

{\bf Figure 4}: {\bf \ }ESR\ absorption line position (corrected from the
sample demagnetization field) as a function of temperature in $%
Nd_{0.5}Ca_{0.5}MnO_{3}$ taken at $95GHz$. The dotted line coresponds to $%
g=1.99$. Insert: ESR absorption spectrum at different temperatures.


\begin{references}
\bibitem{Review}  Y.Tokura, N. Nagaosa, Science {\bf 288}, 462 (2000)

\bibitem{PrCa}  Y. Tomioka{\it \ et al}, Phys. Rev. B {\bf 53}, R1689 (1996)

\bibitem{PRBNdCa}  F. Millange {\it et al}, Phys. Rev. B {\bf 62}, 5619
(2000)

\bibitem{Allodi}  G.\ Allodi {\it et al}, Phys. Rev.\ Lett.\ {\bf 81}, 4736
(1998), G.\ Allodi {\it et al}, Phys. Rev. B. {\bf 61}, 5924 (2000)

\bibitem{Uehara}  M. Uehara {\it et al}, Nature {\bf 399}, 560 (1999)

\bibitem{Franck}  O. Richard {\it et al}, Acta Cryst.\ {\bf A55}, 704 (1999)

\bibitem{phase diagram NdCa}  M. Tokunaga {\it et al}, Phys. Rev.\ B {\bf 57}%
, 5259 (1998) and Phys.\ Rev.\ B {\bf 60}, 6219 (1999)

\bibitem{AMRfred}  F.\ Dupont {\it et al}, App. Magn. Reson.\ {\bf 19}, 485
(2000)

\bibitem{LaCafilm}  F. Dupont {\it et al}, Solid State Com. {\bf 113}, 499
(2000)

\bibitem{LaCadoublespectra}  A.K.\ Srivastava {\it et al}, Solid State Com. 
{\bf 99}, 161 (1996), M.T.\ Causa {\it et al} , Phys. Rev.\ B 58, 3233 (1998)

\bibitem{Chauvet}  O. Chauvet {\it et al}, Phys. Rev. Lett. {\bf 81}, 1102
(1998)
\end{references}
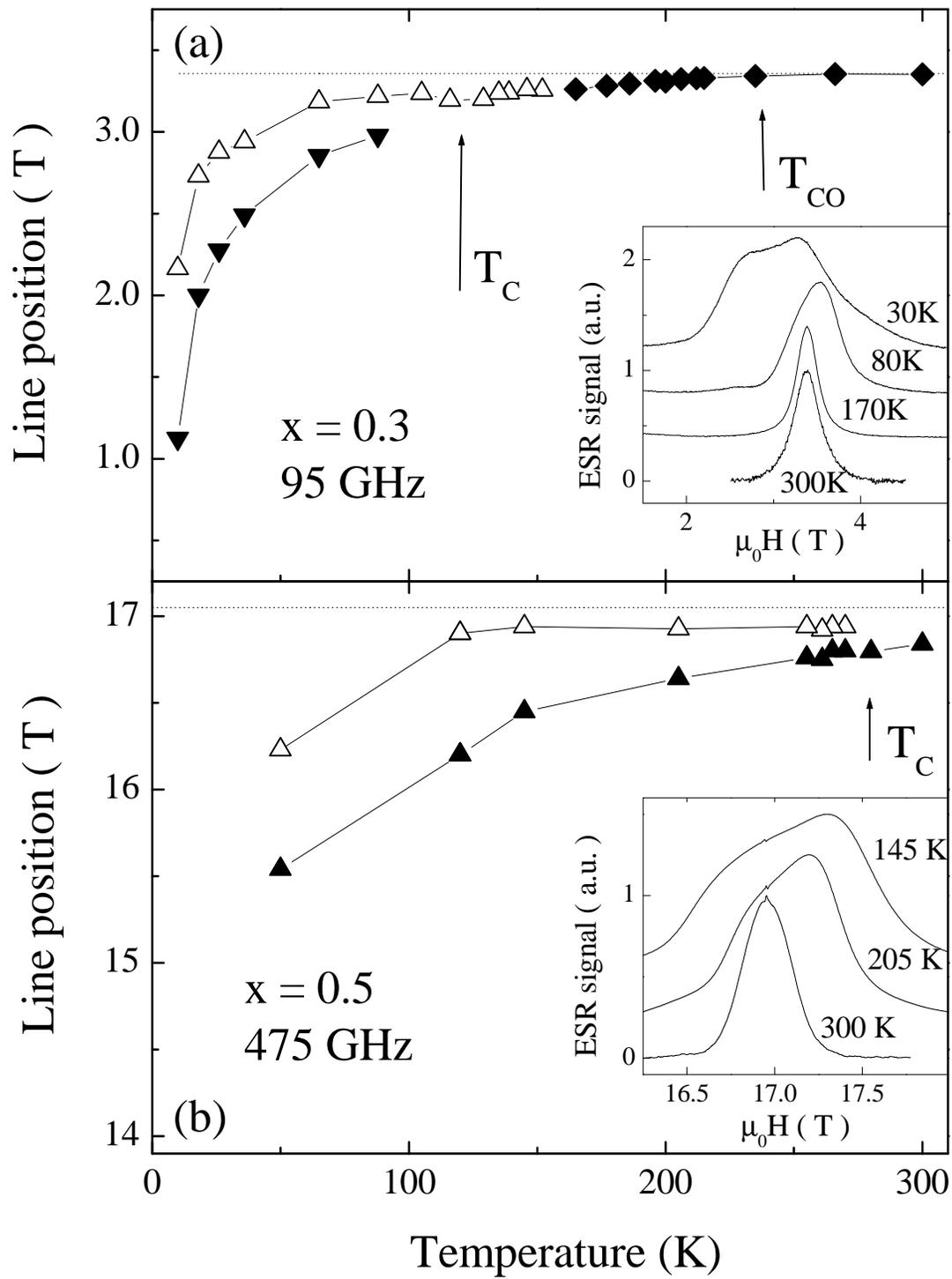
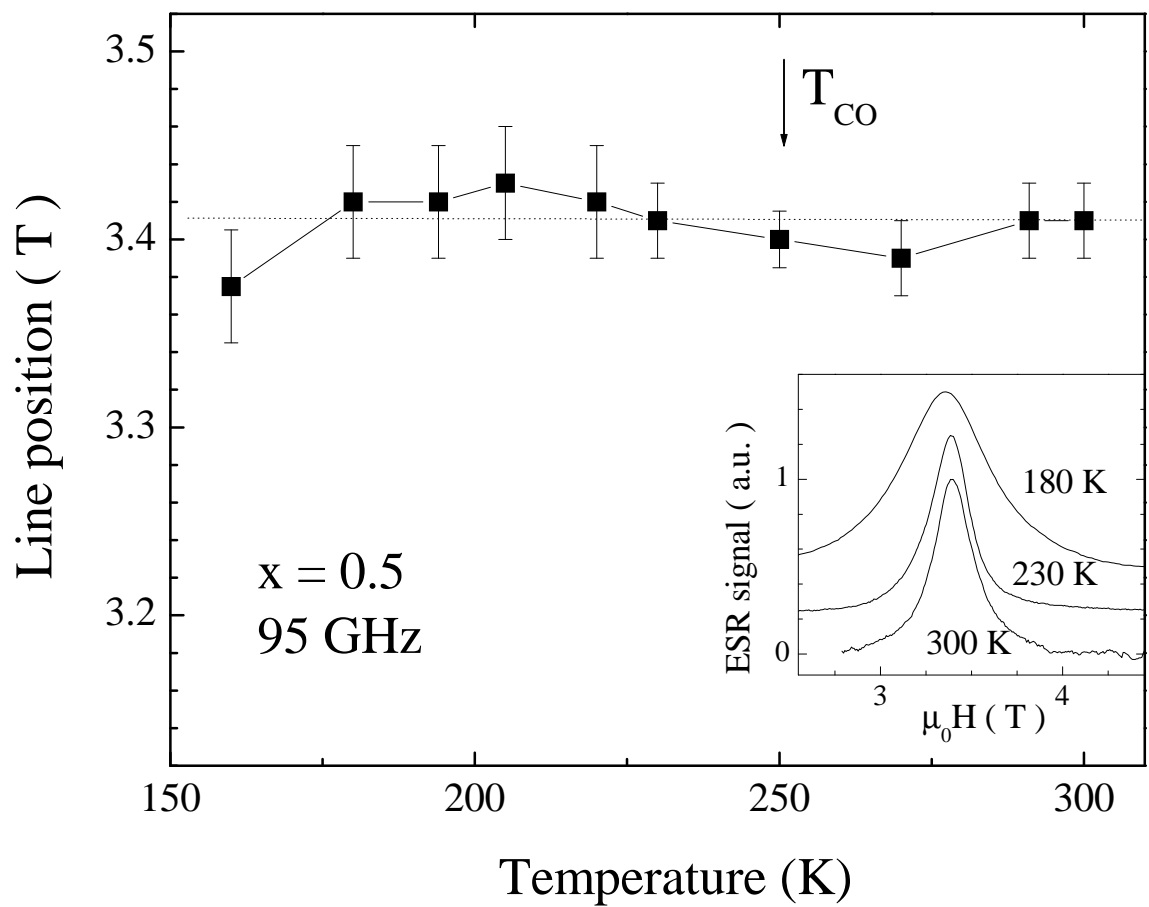
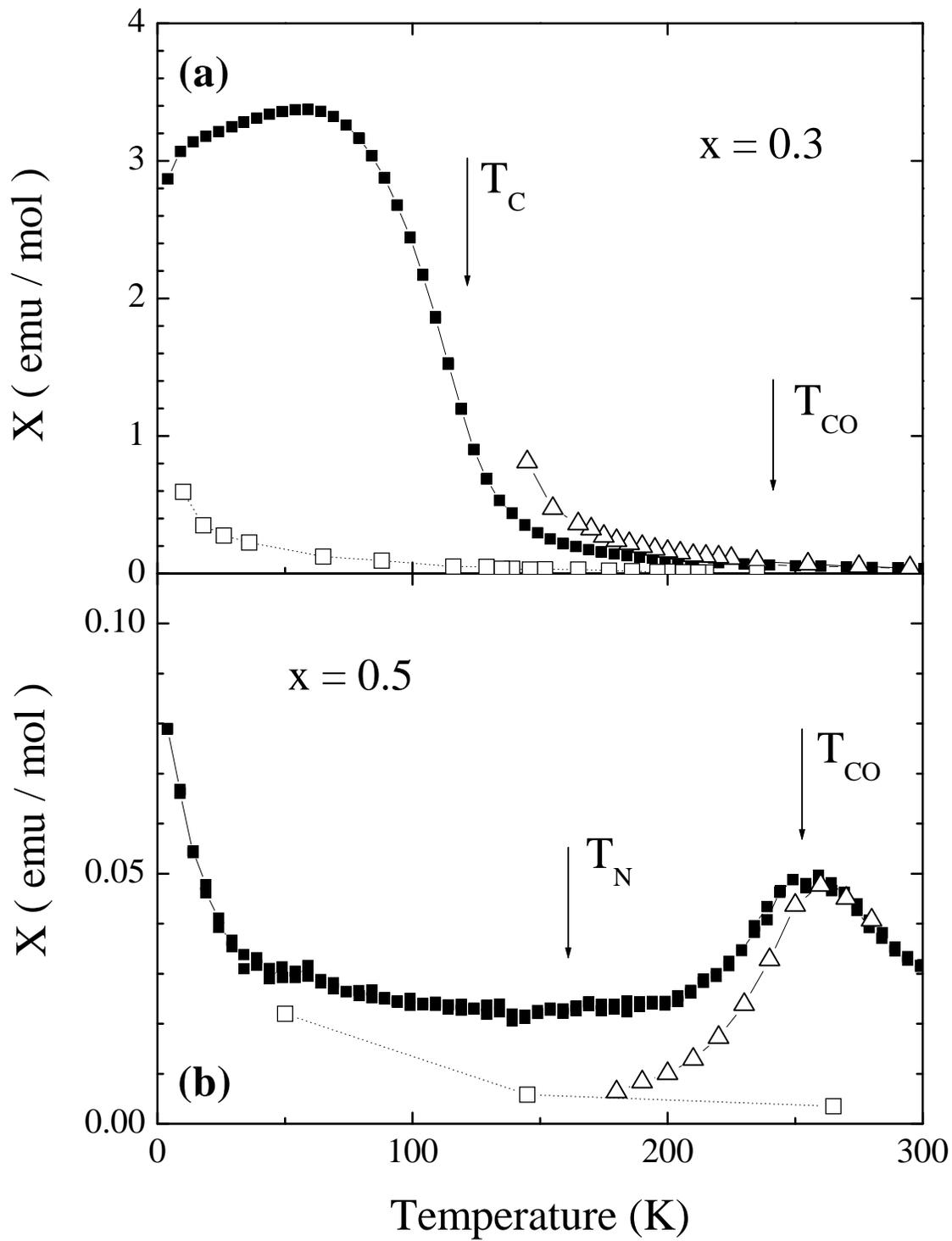
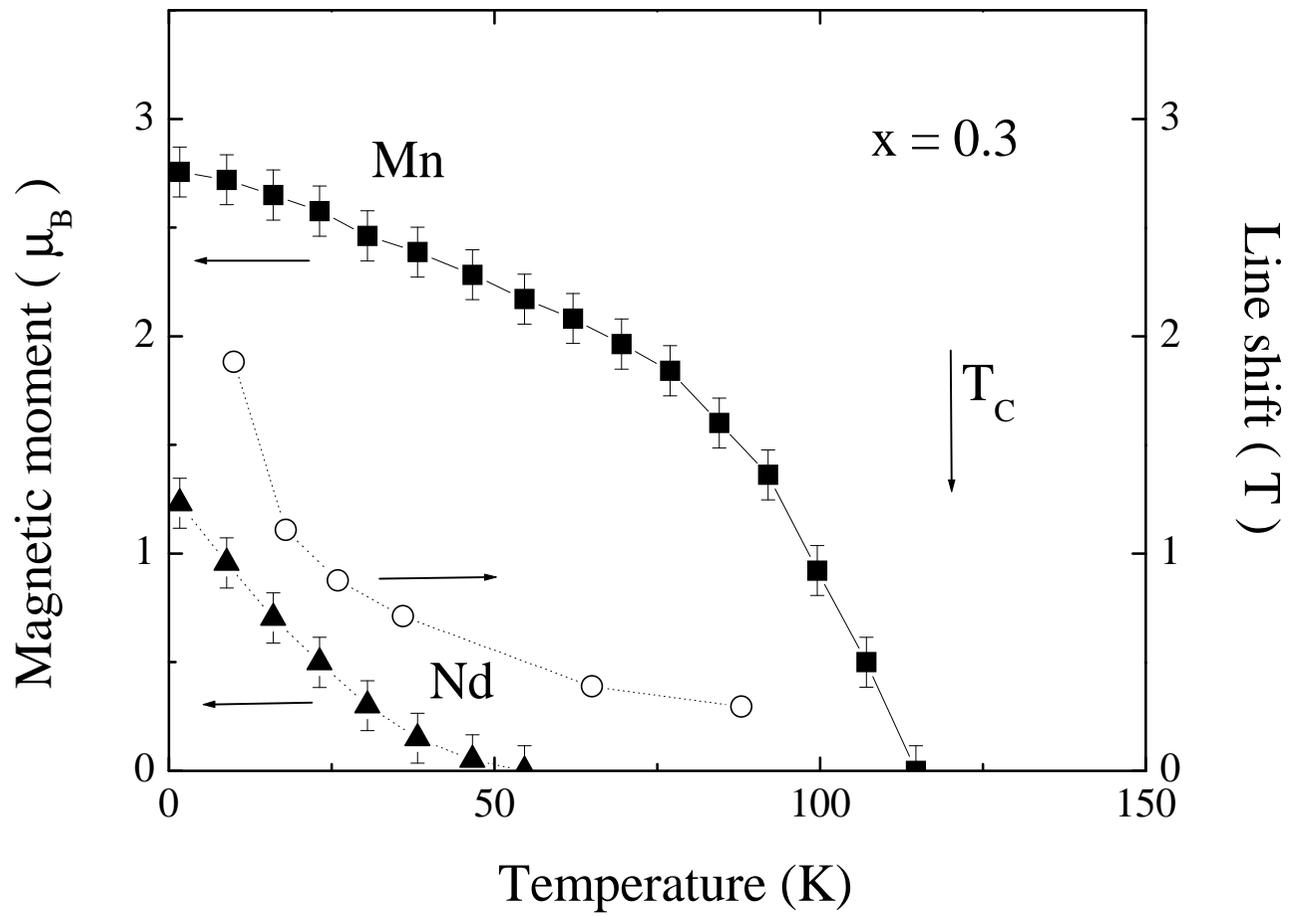
\end{document}